\documentclass{article}
\usepackage{amssymb}
\usepackage{amsmath}
\usepackage{amsfonts}
\usepackage{graphicx}
\usepackage{epstopdf}
\usepackage{placeins}
\usepackage{float}

\begin{document}

\title{{\bf {$X_2$ series of universal quantum dimensions}\vspace{.2cm}}
\author{{\bf M.Y. Avetisyan$^{a,b}$} \ and {\bf R.L. Mkrtchyan$^{b}$}}
\date{ }
}

\maketitle





\begin{center}
$^a$ {\small {\it Yerevan State University, Yerevan, Armenia}}\\
$^b$ {\small {\it Yerevan Physics Institute, Yerevan, Armenia}}\\
\end{center}

\vspace{1cm}

\begin{abstract}
The antisymmetric square of the adjoint representation of any simple Lie algebra is equal to the sum of adjoint and $X_2$ representations. We present universal formulae for quantum dimensions of an arbitrary Cartan power of $X_2$. They are analyzed for singular cases and permuted universal Vogel's parameters. $X_2$ has been the only representation in the decomposition of the square of the adjoint with unknown universal series. Application to universal knot polynomials is discussed. 

MSC classes: 17B20, 17B37, 57M25
\end{abstract}


\section{Introduction}

The universal formulae for simple Lie algebras were first derived by P. Vogel in his Universal Lie Algebra \cite{V0,V}, see also presentation in \cite{R13}. The main aim was to derive the most general weight system for Vassiliev's finite knot invariants. This program met difficulties, however, as a byproduct there appeared the uniform parameterization of simple Lie algebras by the values of Casimir operators on three representations, appearing in decomposition of the symmetric square of the adjoint representations:

\begin{eqnarray}\label{sad}
S^2 \mathfrak{g}=1+Y_2(\alpha)+Y_2(\beta)+Y_2(\gamma)
\end{eqnarray}

One denotes the value of the second Casimir operator on the adjoint representation $\mathfrak{g}$ as $2t$, and parameterizes the values of the same operator on representations in (\ref{sad}) as $4t-2\alpha, 4t-2\beta, 4t-2\gamma$ correspondingly (hence notation of representations in (\ref{sad})). It appears that $\alpha+\beta+\gamma=t$. The values of the parameters for all simple Lie algebras are given in the table \ref{tab:V1}, and in the table \ref{tab:V2} in another form. According to the definitions, the entire theory is invariant with respect to rescaling of the parameters (which corresponds to rescaling of invariant scalar product in algebra), and with respect to the permutation of the universal (=Vogel's) parameters  $\alpha, \beta, \gamma$. So, effectively they belong to a projective plane, which is factorized w.r.t. its homogeneous coordinates, and is called Vogel's plane. 
\begin{table}[ht]
\caption{Vogel's parameters for simple Lie algebras}     \label{tab:V1}

\begin{tabular}{|c|c|c|c|c|c|}
\hline
Root system & Lie algebra  & $\alpha$ & $\beta$ & $\gamma$  & $t=h^\vee$\\   
\hline    
$A_n$ &  $\mathfrak {sl}_{n+1}$     & $-2$ & 2 & $(n+1) $ & $n+1$\\
$B_n$ &   $\mathfrak {so}_{2n+1}$    & $-2$ & 4& $2n-3 $ & $2n-1$\\
$C_n$ & $ \mathfrak {sp}_{2n}$    & $-2$ & 1 & $n+2 $ & $n+1$\\
$D_n$ &   $\mathfrak {so}_{2n}$    & $-2$ & 4 & $2n-4$ & $2n-2$\\
$G_2$ &  $\mathfrak {g}_{2}  $    & $-2$ & $10/3 $& $8/3$ & $4$ \\
$F_4$ & $\mathfrak {f}_{4}  $    & $-2$ & $ 5$& $ 6$ & $9$\\
$E_6$ &  $\mathfrak {e}_{6}  $    & $-2$ & $ 6$& $ 8$ & $12$\\
$E_7$ & $\mathfrak {e}_{7}  $    & $-2$ & $ 8$& $ 12$ & $18$ \\
$E_8$ & $\mathfrak {e}_{8}  $    & $-2$ & $ 12$& $20$ & $30$\\
\hline  
\end{tabular}
\end{table}

\begin{table}[ht] 
 \caption{Vogel's parameters for simple Lie algebras: lines}
\begin{tabular}{|r|r|r|r|r|r|} 
\hline Algebra/Parameters & $\alpha$ &$\beta$  &$\gamma$  & $t$ & Line \\ 
\hline  $\mathfrak {sl}_{N}$  & -2 & 2 & $N$ & $N$ & $\alpha+\beta=0$ \\ 
\hline $\mathfrak {so}_{N}$ & -2  & 4 & $N-4$ & $N-2$ & $ 2\alpha+\beta=0$ \\ 
\hline  $ \mathfrak {sp}_{N}$ & -2  & 1 & $N/2+2$ & $N/2+1$ & $ \alpha +2\beta=0$ \\ 
\hline $Exc(n)$ & $-2$ & $2n+4$  & $n+4$ & $3n+6$ & $\gamma=2(\alpha+\beta)$\\ 
\hline 
\end{tabular}

{For the exceptional 
line $n=-2/3,0,1,2,4,8$ for $\mathfrak {g}_{2}, \mathfrak {so}_{8}, \mathfrak{f}_{4}, \mathfrak{e}_{6}, \mathfrak {e}_{7},\mathfrak {e}_{8} $, 
  respectively.} \label{tab:V2}
\end{table}

As an example of application of this parametrization universal formulae \cite{V0,LM1}, for dimensions of representations from (\ref{sad}) are presented below:

\begin{eqnarray}
\text{dim}\, \mathfrak{g} &=&\frac{(2t-\alpha)(2t-\beta)(2t-\gamma)}{\alpha\beta\gamma} \\
 \label{Y}
 \text{dim} \, Y_2(\alpha)&=& \frac{\left(  2t  - 3\alpha \right) \,\left( \beta - 2t \right) \,\left( \gamma - 2t \right) \,t\,\left( \beta + t \right) \,
      \left( \gamma + t \right) }{\alpha^2\,\left( \alpha - \beta \right) \,\beta\,\left( \alpha - \gamma \right) \,\gamma}
\end{eqnarray}
and  other two (\ref{Y}) representations which are obtained by permutations of the parameters. These are typical universal formulae for dimensions: ratios of products of linear homogeneous functions of universal parameters.

In works of Deligne et al. \cite{Del,DM} one exploits the fact that all exceptional Lie algebras  are located on a line on Vogel's plane, just like $sl(n)$ and $so(n)/sp(n)$ algebras. The idea is that the exceptional line is similar to the ones of special linear and orthogonal algebras, namely, for the exceptional algebras it is also possible to represent the dimensions of all irreps as a ratio of products of linear functions of parameters on the exceptional line. This has been checked for all irreps, appearing in the decomposition of up to the fourth order of the adjoint representation in \cite{Cohen}, and will be compared with our results below. We shall call this "universality on the exceptional line".  Whenever both universal and universal on the exceptional line formulae exist, they coincide.

There are a number of universal formulae for different objects in the theory and applications of simple Lie algebras. E.g. Vogel \cite{V0} found complete decomposition of third power of the adjoint representation in terms of  Universal Lie Algebra, defined by himself, and universal dimension formulae for all representations involved. Landsberg and Manivel \cite{LM1} present a method which allows derivation of certain universal dimension formulae for simple Lie algebras and derive those for Cartan powers of the adjoint, $Y_2(.)$, and their Cartan products. Westbury \cite{W3} found a universal formula for quantum dimension of the adjoint representation. Sergeev, Veselov and one of the present authors derived \cite{MSV} a universal formula for generating function for the eigenvalues of higher Casimir operators on the adjoint representation. 

In subsequent works applications to physics were developed, particularly the universality of the partition function of Chern-Simons theory on a sphere \cite{MV,M13,KhM16-1}, and its connection with q-dimension of $k\Lambda_0$ representation of affine Kac-Moody algebras \cite{M17} were shown, the universal knot polynomials for 2- and 3-strand torus knots \cite{W15,MMM,MM,M16QD} were calculated.

Another application of universal formulae is the derivation of non-perturbative corrections to Gopakumar-Vafa partition function \cite{KM,KS} by gauge/string duality from the universal partition function of Chern-Simons theory. This shows the relevance of the "analytical continuation" of the universal formulae from the points of Vogel's table (\ref{tab:V1}) to the entire Vogel's plane.   

A completely different direction of development - the Diophantine classification of simple Lie algebras \cite{M16} and its connection with  McKay correspondence, \cite{KhM16-2} is also worth mentioning.

In the present paper our goal is to derive universal formulae for quantum dimensions for an arbitrary Cartan power of the  $X_2$ representation, appearing in the following decomposition

\begin{eqnarray}
\wedge^2 \mathfrak{g}=\mathfrak{g}+X_2
\end{eqnarray}

$k$-th Cartan power of a representation with highest weight $\lambda$ is that with highest weight  $k\lambda$. Note that for  $sl(n)$ algebras $X_2$ is not an irreducible representation, until one considers the Lie algebra's semidirect product with the automorphism group of its Dynkin diagram (instead of the algebra itself), as suggested and implemented in \cite{Del,DM,Cohen} for the exceptional algebras. Particularly, in  $sl(n)$  case one has $Z_2$ as an automorphism group and $X_2$ is the sum of representations with highest weights $2\omega_1+\omega_{n-2}$ and $\omega_2+2\omega_{n-1}$. Its Cartan power we consider to be the sum of Cartan powers of these two representations. More generally, any irrep of simple Lie algebras below are considered to be extended by the automorphism group of their Dynkin diagram. We shall see, that universal formulae give answers for irreps of such extended Lie algebras, i.e. if there appears an irrep which is not invariant under automorphism, then it appears in the sum with his automorphism-transformed version(s), so that the invariance is recovered. 

For $k=1$ the universal quantum dimension of $X_2$ is given in \cite{D13}:

\begin{eqnarray} \nonumber
D_Q^{X_2}=\frac{\sinh\left(\frac{x}{4}(2t-\alpha)\right)\sinh\left(\frac{x}{4}
	(2t-\beta)\right)\sinh\left(\frac{x}{4}(2t-\gamma)\right)}{\sinh\left(\frac{\alpha x}{4}\right)\sinh\left(\frac{\beta x}{4}\right)\sinh\left(\frac{\gamma x}{4}\right)}
\times \\ \nonumber
\frac{\sinh\left(\frac{x}{4}(t+\alpha)\right)\sinh\left(\frac{x}{4}(t+\beta)\right)\sinh\left(\frac{x}{4}(t+\gamma)\right)}{\sinh\left(\frac{\alpha x}{2}\right)\sinh\left(\frac{\beta x}{2}\right)\sinh\left(\frac{\gamma x}{2}\right)} \times \\ \frac{\sinh\left(\frac{x}{2}(t-\alpha)\right)\sinh\left(\frac{x}{2}(t-\beta)\right)\sinh\left(\frac{x}{2}(t-\gamma)\right)}{\sinh\left(\frac{x}{4}(t-\alpha)\right)\sinh\left(\frac{x}{4}(t-b)\right)\sinh\left(\frac{x}{4}(t-\gamma)\right)} \label{k=1}
\end{eqnarray}

The proof and discussion of the properties of this and the general formula for $k>1$ is given further in the paper.  

Up to now, $X_2$ remains the only representation from the square of the adjoint representation, which has no universal formulae for (quantum) dimensions of its Cartan powers. For powers of other representations, i.e. $Y_2(.)$, both usual and quantum dimensions are given in \cite{LM1,MMM,M17}.

In the next sections we outline the way in which we derive the final formula for quantum dimension of $k$-th Cartan power of $X_2$.The derivation is actually an "educated guess" based on the data from some exceptional algebras and knowledge of existing universal formulae. So, it needs to be proven, which is carried out in the Appendix.The universal formulae make sense for permutations of Vogel's parameters, and in Section \ref{sing} we consider our formula for quantum dimensions of powers of $X_2$ with permuted parameters. We show that it gives quantum dimensions for some other representations of corresponding simple Lie algebras (times automorphisms group). In Conclusion we discuss the applications of present results to the universal knot polynomials and representation theory.

\section{Technique}

There is no regular way of obtaining universal formulae (and their very existence is not guaranteed). Vogel's approach gave unique answers for dimensions, but it is based on the calculation with ring $\Lambda$, which appears to have \cite{V} divisors of zero, so that approach is not self-consistent if one does not  handle that issue carefully. In \cite{LM1} (and in the present work) the restricted definition of universal formulae is adopted, as defined above - they have to give correct answers for true simple Lie algebras in the corresponding points of Vogel's table \ref{tab:V1}. 

That allows one to use the Weyl formula for characters, restricted to the Weyl line, i.e. for quantum dimensions (see e.g. \cite{DiF}, 13.170): 

\begin{eqnarray}\label{W}
D_Q^\lambda= 
\chi_{\lambda}(x\rho)= \prod_{\mu >0} \frac{\sinh(\frac{x}{2}(\mu,\lambda+\rho))}{\sinh(\frac{x}{2}(\mu,\rho))}
\end{eqnarray}

where $\lambda$ is the highest root of the given irreducible representation, $\rho$ is the Weyl's vector, the sum of the fundamental weights.The usual dimensions are obtained in the $x \rightarrow 0$ limit of the quantum ones. Both sides of this formula are invariant w.r.t. the simultaneous rescaling (in "opposite directions") of the scalar product in algebra and the parameter $x$.The automorphism of the Dynkin diagram leads to the equality of quantum dimensions for representations with highest weights connected by automorphism.

Evidently, only the roots with non-zero scalar product with  $\lambda$ contribute. So, one has to express the scalar product of such roots with  $\lambda$ and $\rho$ in terms of the universal parameters, and that has to be done in a uniform way for all simple Lie algebras. Then one may hope to get a universal expression for $D_\lambda$. 

To describe the technique, consider, e.g. the case of $\lambda=\theta$, the highest weight of the adjoint representation. As it is shown in \cite{LM1}, the values of scalar product of roots with $\theta$ are either $2$ (for root $\theta$ itself) or $1$. These last roots  can be organized into three "segments" (see definition below) with unit spacing of $(\rho,\alpha)$ (we normalize the scalar product as in \cite{LM1} and table \ref{tab:V1} by $\alpha=-2$), which we present below for $E_7$ as an example:

\begin{table}[h]
\caption{Height $ht=(\rho,\mu)$ and $n_{ht}$ for all roots $\mu$ with $(\theta,\mu)=1$ for $E_7$}
	\centering
		\begin{tabular}{|l|r|r|r|r|r|r|r|r|r|r|r|r|r|r|r|r|}
		\hline
		$ht$&1&2&3&4&5&6&7&8&9&10&11&12&13&14&15&16\\
		\hline
				$n_{ht}$ &1&1&1&2&2&3&3&3&3&3&3&2&2&1&1&1\\
				\hline
		\end{tabular}
		\label{tab:E7-black}
		\end{table}

where in the first line there are the values of scalar products with $\rho$, i.e. the heights $ht$ of roots, in the second -  $n_{ht}$ - the number of roots on that height (remember we consider the roots $\mu$ with $(\mu,\theta)$, only). So, we see, that roots with $(\theta,\mu)=1$ can be organized into three sets of roots, which we shall call "segments of roots", or simply segments. Segment of roots is the finite number of roots with equidistant values of heights including exactly one root for any given height from that equidistant sequence of heights. The first, the longest segment, has length $t-2=16$, with heights from $1$ to $16$, the second is in the center of the first, is of length $\gamma-2=10$ (we order universal parameters as $\gamma \geq \beta >-2)$, and the third segment, again in the center of the first (and the second) segments, has length $\beta-2$. The same pattern of segments is observed for most of the simple Lie algebras. 

With this data it is easy to obtain universal formulae for dimensions \cite{LM1} and quantum dimensions \cite{M17} for $k$-th Cartan power of the adjoint representation. Namely, numerators and denominators of consecutive roots of the given segment of roots cancel (\ref{D}), so for each segment there remains a number of the first denominators and the same number of the last numerators, which finally lead to the universal formulae.

These results have been proven in \cite{LM1} partially by "general" considerations, restricted, however, to the algebras of the rank at least three, and partially by case by case considerations for each algebra separately.

The description above reflects the advantage of the approach - possibility of using Weyl formula, as a basis of calculations, and shortcomings, which come from the use of very restricted sets of truly existing simple Lie algebras, see more on that below. Particularly, one can add an arbitrary polynomial to the results, which accepts zero values on the lines of the simple Lie algebras (tables \ref{tab:V1}, \ref{tab:V2}). Such "minimal" symmetric polynomial can be easily written:

\begin{eqnarray}
(\alpha+\beta)(\beta+\gamma)(\gamma+\alpha)(2\alpha+\beta)(2\beta+\alpha)(2\alpha+\gamma)(2\gamma+\alpha)\cdot \\\nonumber
(2\beta+\gamma)(2\gamma+\beta)(2\gamma+2\alpha-\beta)(2\gamma+2\beta-\alpha)(2\alpha+2\beta-\gamma)
\end{eqnarray}

However, one can require that, first, the formula should be presented as a ratio of products of linear functions over universal parameters (and not the sum of such expressions), and, second, that Deligne hypothesis \cite{D13} should be satisfied. Deligne assumes that the standard relations of characters (recall that quantum dimensions are characters on the Weyl line) namely, the product of characters of two representations is equal to the sum of characters of their decomposition, should be satisfied on the entire Vogel's plane (and not on the points of Vogel's table, only). Deligne's hypothesis is checked in some cases \cite{M16QD}, particularly for symmetric cube of the adjoint representation. At this time it is not known whether it is possible to satisfy one or both of these requirements, as well as the very existence of universal formulae is not guaranteed. So, we do not worry on this problem further in this paper, and present the new universal formulae in the natural way we found them. 

So, below we use this approach to obtain the universal formulae for quantum dimensions of $k$-th Cartan powers of  $X_2$ representation.

In the next section we present data for $E_n$ algebras and try to rewrite them in the universal form. It appears that it is not sufficient for derivation of general formula, due to the ambiguities of rewriting the answers in the universal form. We use two additional ideas: first is that the results should not be singular for $sl(n)$ algebra, and, second, that the answer should be invariant w.r.t. the permutation of two parameters. In that way we obtain the final formula (\ref{D}) below. All this, however, does not combine into formal derivation and altogether should be considered as an educated guess. The formal proof is carried out in the Appendix, for all algebras. We nevertheless outline these steps to show how we came to the final, sufficiently complicated formula.The development of a general method for derivation of universal formulae still remains an open problem.

\section{$E_n$ data}

It appears that $E_n$ are the only algebras, which can hint on a universal form of non-trivial contributions to the Weyl formula (\ref{D}) for $X_2$ representation, the highest weight of which we denote $\lambda$. So below we present relevant roots and their contributions. Everywhere we use Dynkin's labeling of Dynkin diagram. 

\subsection{$E_8$}

Dimension of $E_8$=248, number of positive roots $|\Delta_+|=120$, Vogel's parameters $(\alpha,\beta,\gamma)=(-2,12,20)$.
For $E_8$ the highest weight of $X_2$ is $\lambda=\omega_6$, in Dynkin's labeling of roots.

The number of positive roots $\mu$ with $(\lambda,\mu)=0$ is $1+|\Delta_+|_{E_6}=1+36=37$. The number of positive roots $\mu$ with $(\lambda,\mu)=1$ is 54 and is given in table \ref{tab:E8-1} with corresponding numbers $n_{ht}$ of roots with given scalar product with $\rho$.

\FloatBarrier
\begin{table}[ht]
\caption{Number $n_{ht}$ vs height $ht=(\rho,\mu)$ for roots $\mu$ with $(\lambda,\mu)=1$ for $E_8$}
	\centering
		\begin{tabular}{|l|r|r|r|r|r|r|r|r|r|r|r|r|r|r|r|r|r|r|}
		\hline
		$ht$&1&2&3&4&5&6&7&8&9&10&11&12&13&14&15&16&17&18\\
		\hline
		$n_{ht}$ &1&2&2&2&3&4&4&4&5&5&4&4&4&3&2&2&2&1 \\
		\hline
		\end{tabular}
		\label{tab:E8-1}
		\end{table}
\FloatBarrier

So, here we have 5 segments of roots. The number of positive roots $\mu$ with $(\lambda,\mu)=2$ is 27 and is given in table \ref{tab:E8-2} with numbers $n_{ht}$.

\FloatBarrier
\begin{table}[h]	
\caption{Number $n_{ht}$ vs height $ht=(\rho,\mu)$ for roots $\mu$ with $(\lambda,\mu)=2$ for $E_8$}
	\centering
		\begin{tabular}{|l|r|r|r|r|r|r|r|r|r|r|r|r|r|r|r|r|r|}
		\hline
		$ht$&11&12&13&14&15&16&17&18&19&20&21&22&23&24&25&26&27 \\
		\hline
		$n_{ht}$ &1&1&1&1&2&2&2&2&3&2&2&2&2&1&1&1&1 \\
		\hline
		\end{tabular}	
		\label{tab:E8-2}
		\end{table}
\FloatBarrier

Here we have 3 segments of roots. The number of positive roots $\mu$ with $(\lambda,\mu)=3$ is 2 and is given in table \ref{tab:E8-3}.

\FloatBarrier
\begin{table}[h]
\caption{Number $n_{ht}$ vs height $ht=(\rho,\mu)$ for roots $\mu$ with $(\lambda,\mu)=3$ for $E_8$}
	\centering
		\begin{tabular}{|l|r|r|}
		\hline
		$ht$&28&29 \\
		\hline
		$n_{ht}$ &1&1 \\
		\hline
		\end{tabular}
		\label{tab:E8-3}
		\end{table}
\FloatBarrier

Now we have 1 segment, consisting of two roots. Check of the total number of positive roots: 37+54+27+2=120.

\subsection{$E_7$}

Dimension $E_7$=133, number of positive roots $|\Delta_+|=63$, Vogel's parameters $(\alpha,\beta,\gamma,t)=(-2,8,12,18)$. For $E_7$ $\lambda=\omega_2$, in Dynkin's labeling of roots. 

The number of positive roots $\mu$ with $(\lambda,\mu)=0$ is $1+|\Delta_+|_{A_5}=1+15=16$.
 The number of positive roots $\mu$ with $(\lambda,\mu)=1$ is 30 and is given in table \ref{tab:E7-1}.

\FloatBarrier
\begin{table}[h]
\caption{Number $n_{ht}$ vs height $ht=(\rho,\mu)$ for roots $\mu$ with $(\lambda,\mu)=1$ for $E_7$}
	\centering
		\begin{tabular}{|l|r|r|r|r|r|r|r|r|r|r|}
		\hline
		$ht$&1&2&3&4&5&6&7&8&9&10\\
		\hline
		$n_{ht}$ &1&2&3&4&5&5&4&3&2&1 \\
		\hline
		\end{tabular}
		\label{tab:E7-1}
		\end{table}
\FloatBarrier

So, here we have 5 segments of roots. The number of positive roots $\mu$ with $(\lambda,\mu)=2$ is 15 and is given in table \ref{tab:E7-2}.

\FloatBarrier
\begin{table}[h]
\caption{Number $n_{ht}$ vs height $ht=(\rho,\mu)$ for roots $\mu$ with $(\lambda,\mu)=2$ for $E_7$}
	\centering
		\begin{tabular}{|l|r|r|r|r|r|r|r|r|r|}
		\hline
		$ht$&7&8&9&10&11&12&13&14&15 \\
		\hline
		$n_{ht}$ &1&1&2&2&3&2&2&1&1 \\
		\hline
		\end{tabular}
		\label{tab:E7-2}
		\end{table}
\FloatBarrier

Here we have 3 segments of roots. The number of positive roots $\mu$ with $(\lambda,\mu)=3$ is 2 and is given in table \ref{tab:E7-3}.

\FloatBarrier
\begin{table}[h]
\caption{Number $n_{ht}$ vs height $ht=(\rho,\mu)$ for roots $\mu$ with $(\lambda,\mu)=3$ for $E_7$}
	\centering
		\begin{tabular}{|l|r|r|}
		\hline
		$ht$&16&17 \\
		\hline
		$n_{ht}$ &1&1 \\
		\hline
		\end{tabular}
		\label{tab:E7-3}
		\end{table}
\FloatBarrier

Now we have 1 segment of roots.  Check of the total number of positive roots: 16+30+15+2=63.

\subsection{$E_6$}

dim$E_6$=78, $|\Delta_+|=36$, $(\alpha,\beta,\gamma,t)=(-2,6,8,12)$. For $E_6$ $\lambda=\omega_3$, in Dynkin's labeling of roots.

The number of positive roots $\mu$ with $(\lambda,\mu)=0$ is $7$. The number of positive roots $\mu$ with $(\lambda,\mu)=1$ is 18 and is given in table \ref{tab:E6-1} together with numbers $n_{ht}$.

\FloatBarrier
\begin{table}[h]
\caption{Number $n_{ht}$ vs height $ht=(\rho,\mu)$ for roots $\mu$ with $(\lambda,\mu)=1$ for $E_6$}
	\centering
		\begin{tabular}{|l|r|r|r|r|r|r|}
		\hline
		$ht$&1&2&3&4&5&6\\
		\hline
		$n_{ht}$ &1&3&5&5&3&1 \\
		\hline
		\end{tabular}
		\label{tab:E6-1}
		\end{table}
\FloatBarrier

So, here we have 5 segments of roots, i.e. sequences with unit distance between consecutive roots. The number of positive roots $\mu$ with $(\lambda,\mu)=2$ is 9 and is given in table \ref{tab:E6-2}.
\FloatBarrier
\begin{table}[h]
\caption{Number $n_{ht}$ vs height $ht=(\rho,\mu)$ for roots $\mu$ with $(\lambda,\mu)=2$ for $E_6$}
	\centering
		\begin{tabular}{|l|r|r|r|r|r|}
		\hline
		$ht$&5&6&7&8&9 \\
		\hline
		$n_{ht}$ &1&2&3&2&1 \\
		\hline
		\end{tabular}
		\label{tab:E6-2}
		\end{table}
\FloatBarrier
Here we have 3 segments of roots.  The number of positive roots $\mu$ with $(\lambda,\mu)=3$ is 2 and is given in table \ref{tab:E6-3}.

\FloatBarrier
\begin{table}[h]
\caption{Number $n_{ht}$ vs height $ht=(\rho,\mu)$ for roots $\mu$ with $(\lambda,\mu)=3$ for $E_6$}
	\centering
		\begin{tabular}{|l|r|r|}
		\hline
		$ht$&10&11 \\
		\hline
		$n_{ht}$ &1&1 \\
		\hline
		\end{tabular}
		\label{tab:E6-3}
		\end{table}
\FloatBarrier
So, here we have 1 segment of roots. Check of the total number of positive roots: 7+18+9+2=36.

\section{Quantum dimensions}

Now we calculate the contributions of roots with $(\lambda,\mu)\neq 0$ in the Weyl formula for quantum dimension. 

The contribution of roots with $(\lambda,\mu)=3$ comes from two roots of heights $t-1,t-2$ (recall the normalization $\alpha=-2$):  

\begin{eqnarray} 
L_3=\frac{\text{sinh}\left(\frac{x}{2}(t+1)\right)\text{sinh}\left(\frac{x}{2}(t+2)\right)}{\text{sinh}\left(\frac{x}{2}(t-2)\right)\text{sinh}\left(\frac{x}{2}(t-1)\right)}
\end{eqnarray} 

Due to the rescaling invariance, mentioned after (\ref{W}), we can recover the parameter $\alpha$ in this formula in explicit form by substitution  
\begin{eqnarray}
\beta \rightarrow -2\beta/\alpha, \gamma \rightarrow -2\gamma/\alpha, t \rightarrow -2t/\alpha,  x\rightarrow -x\alpha/2
\end{eqnarray}
Then $L_3$ accepts the form
\begin{eqnarray} 
L_{3}=\frac{\sinh
\left(\frac{x}{4}
(3 \alpha (k-1)-2 (\beta+\gamma))\right) \sinh
\left(\frac{x}{4} (\alpha (3
k-4)-2
(\beta+\gamma))\right)}{\sinh\left(\frac{x}{2} (2
\alpha+\beta+\gamma)\right)
\sinh\left(\frac{x}{4}  
(3 \alpha+2 (\beta+\gamma))\right)}
\end{eqnarray}

Below we skip the intermediate formulae in normalization $\alpha=-2$, and present the final ones with explicit $\alpha$ recovered.

Next consider roots with $(\lambda,\mu)=2$. There are three segments, the first (longest) one starts at height $\beta-1$ and ends at height $t-3$, its contribution in the Weyl formula is 

\begin{eqnarray} 
L_{21}=\prod _{i=1}^{2 k} \frac{\sinh \left(\frac{1}{4}
x (\alpha (i-5)-2
(\beta+\gamma))\right)}{\sinh\left(\frac{1}{4} x
(\alpha
(i-2)-2
\beta)\right)}
\end{eqnarray}

The second segment starts at height $t/2$ and ends at height $(t+\gamma-4)/2$, the contribution is 
\begin{eqnarray} 
L_{22}=\prod _{i=1}^{2 k} \frac{\sinh \left(\frac{1}{4} x
(-\alpha
(i-3)+\beta+2 \gamma)\right)}{\sinh\left(\frac{1}{4} x
(-\alpha
(i-2)+\beta+\gamma)\right)}
\end{eqnarray} 

The third segment includes one root at height $(\gamma+2\beta-6)/2$ and it's contribution is
\begin{eqnarray} 
L_{23}=\frac{\sinh
\left(\frac{1}{4} x (\alpha (3-2 k)+2
\beta+\gamma)\right)}{\sinh\left(\frac{1}{4} x (3 \alpha+2
\beta+\gamma)\right)} 
\end{eqnarray}

Next are the roots with $(\lambda,\mu)=1$. There are five segments, the first (longest) one starts at height $1$ and ends at height $(\gamma+2\beta-8)/2$, its contribution in the Weyl formula is 

\begin{equation}
L_{11}=\prod _{i=1}^k \frac{\sinh
\left(\frac{1}{4} x (-\alpha (i-4)+2
\beta+\gamma)\right)}{\sinh\left(\frac{\alpha i x}{4}\right)}
\end{equation}

The second segment starts at height $2$ and ends at height $\gamma-2$, contributing 

\begin{eqnarray} 
L_{12}=\prod _{i=1}^k \frac{\sinh
\left(\frac{1}{4} x (\alpha (i-3)-2 \gamma )\right)}{\sinh
\left(\frac{1}{4} \alpha (i+1) x\right)}
\end{eqnarray} 

The third segment starts at height $(\beta-2)/2$ and ends at $(\gamma+\beta-4)/2$, contributing

\begin{eqnarray} \label{L13}
L_{13}=\prod _{i=1}^k \frac{\sinh \left(\frac{1}{4} x
(-\alpha
(i-2)+\beta+\gamma)\right)}{\sinh\left(\frac{1}{4} x (\beta-
\alpha
(i-2))\right)}
\end{eqnarray} 

The fourth segment is similar to the third one, but shorter by one element on each end:

\begin{eqnarray} 
L_{14}=\prod _{i=1}^k\frac{\sinh \left(\frac{1}{4} x
(-\alpha (i-3)+\beta+\gamma)\right)}
{\sinh \left(\frac{1}{4} x (\alpha
(-i)+\alpha+\beta)\right)} 
\end{eqnarray} 

The fifth segment consists of two roots, starting at height $(\gamma-2)/2$, and contribution will be

\begin{eqnarray}\label{L15exp}
\frac{\text{sinh}\left(\frac{x}{2}(\beta-3+k)\right)}{\text{sinh}\left(\frac{x}{2}\left(\frac{\gamma-2}{2}\right)\right)}\frac{\text{sinh}\left(\frac{x}{2}(\beta-2+k)\right)}{\text{sinh}\left(\frac{x}{2}\left(\frac{\gamma-2}{2}+1\right)\right)}
\end{eqnarray}

This contribution is appropriate at $k=1$, in a sense that all contributions together - the product of all $L$-s - form the corrects answer (\ref{k=1}). However, for $k>1$ and for $sl(n)$ algebras (i.e. on the line $\alpha+\beta=0$) one loses the zero of (\ref{L15exp}) on that line which at $k=1$ cancels out with the zero in denominator of (\ref{L13}), also on the same line. So,
in analogy with other contributions above, we simply change this contribution to other one, namely $L_{15}$, written below. It cancels mentioned singularity for an arbitrary $k$, coincides with (\ref{L15exp}) on the exceptional line $\gamma=2(\alpha+\beta)$, but differs in other points:

\begin{eqnarray} \label{L15}
L_{15}=\prod _{i=1}^k\frac{\sinh \left(\frac{1}{4} x (\alpha
(i-3)-2
\beta)\right) \sinh
\left(\frac{1}{4} x (\alpha (i-2)-2 \beta)\right)
}{\sinh\left(\frac{1}{4}
x (\gamma-\alpha
(i-2))\right) \sinh\left(\frac{1}{4} x (\alpha
(-i)+\alpha+\gamma)\right)} 
\end{eqnarray}

However, this is not the end of the story. We expect that our final formula should be invariant under switch of the $\beta$ and
$\gamma$ parameters, in analogy with the universal formula (\ref{sad}) for  $Y_2(\alpha)$ . So we add a new multiplier, which  in some "minimal" way symmetrizes the product of all $L_?$ multipliers above w.r.t. the switch $\beta \leftrightarrow \gamma$:
\begin{eqnarray} \label{Lc}
L_{corr}=\prod _{i=1}^k\frac{\sinh \left(\frac{1}{4} x (\alpha
(-(i+k-4))+2
\beta+\gamma)\right)}{\sinh\left(\frac{1}{4} x (\alpha
(i+k-2)-2
\gamma)\right)}
\end{eqnarray}

Finally, our main result is 

{\bf Proposition.}

{\it The function 
\begin{eqnarray} \label{D}
X_2(x,k,\alpha,\beta,\gamma)\equiv X_2(k,\alpha)= L_3L_{21}L_{22}L_{23}L_{11}L_{12}L_{13}L_{14}L_{15}L_{corr}
\end{eqnarray}
is equal, besides exceptions, to the quantum dimensions of $k$-th Cartan power of above defined $X_2$ representation for any given simple Lie algebra on corresponding point of Vogel's table \ref{tab:V1}. Exceptions are: $sp(2n)$, for which the formula gives the quantum dimensions of $X_2$ at $k=1$, and zero otherwise, and the $B_2$ algebra. Exact details are given in the tables \ref{tab:x2acl} and \ref{tab:x2aex}.}

The case by case proof of {\bf Proposition} is given in the Appendix.

 \FloatBarrier
 \begin{table}[h]
   \caption{$X_2(k,\alpha)$ for classical algebras}
     \begin{tabular}{|c|p{2cm}|p{2cm}|c|}
     \hline
          $k$&1&2&$\geq3$ \\
          \hline
        $A_1$&0&0&0\\  
        \hline
        $A_2$&$3\omega_1\oplus 3\omega_2$&$
        6\omega_1\oplus 6\omega_2$& $3k\omega_1\oplus 3k\omega_2$
        \\  
        \hline
         $A_n,n\geq3$&$(2\omega_1+\omega_{n-1})\oplus 
       (\omega_2+2\omega_{n})$&$2(2\omega_1+\omega_{n-1})\oplus
       2(\omega_2+2\omega_{n})$&$k(2\omega_1+\omega_{n-1})\oplus
       k(\omega_2+2\omega_{n})$\\
       \hline
        $B_2$&$\omega_1+2\omega_2$&0&0\\  
        \hline
        $B_3$&$\omega_1+2\omega_3$&$2\omega_1+4\omega_3$
        &
        $k(\omega_1+2\omega_3)$\\  
        \hline
        $B_n,n\geq4$&$\omega_1+\omega_3$&$2(\omega_1+
       \omega_3)$&$k(\omega_1+
       \omega_3)$\\
       \hline
        $C_n,n\geq3$&$2\omega_1+\omega_2$&0&0\\
       \hline
        $D_4$&$\omega_1+\omega_3+\omega_4$&$
        2(\omega_1+\omega_3+\omega_4)$&
        $k(\omega_1+\omega_3+\omega_4)$\\
        \hline
        $D_n,n\geq5$&$\omega_1+\omega_3$&$2(\omega_1+
       \omega_3)$&$k(\omega_1+ \omega_3)$\\
       \hline
       \end{tabular}
       \label{tab:x2acl}
         \end{table}     
         \FloatBarrier

 \FloatBarrier
 \begin{table}[h]
 \caption{$X_2(k,\alpha)$ for exceptional algebras}
     \begin{tabular}{|c|c|c|}
     \hline
          $k$&1&$\geq2$\\
          \hline
       $G_2$&$3\omega_1$&$3k\omega_1$\\
       \hline
       $F_4$&$\omega_2$&$k\omega_2$\\
       \hline
       $E_6$&$\omega_3$&$k\omega_3$\\
       \hline
       $E_7$&$\omega_2$&$k\omega_2$\\
       \hline
    $E_8$&$\omega_6$&$k\omega_6$\\
       \hline
       $D_4$&$\omega_1+\omega_3+\omega_4$&
       $k(\omega_1+\omega_3+\omega_4)$\\
       \hline
       \end{tabular}
       \label{tab:x2aex}
    \end{table}   
\FloatBarrier

{\bf Remark 1, on $sl(n)$ case.} In the case of $sl(n)$ line  denominator of $L_{13}$ and numerator of $L_{15}$ both contain a zero multiplier, which however cancel out, i.e. one can continuously extend $X_2(k,\alpha)$ function on that line. In more detail: for the $\alpha+\beta=0$ line the mentioned fraction is 
\begin{eqnarray}
\frac{\sinh((2\beta+2\alpha)x/4)}{\sinh((\beta+\alpha)x/4)} 
\end{eqnarray}

and evidently tends to $2$ in the limit $\alpha+\beta \rightarrow 0$ independent on the direction of approaching the given point on the line on Vogel's plane. Of course, one can simply substitute the expression 
\begin{eqnarray}
\frac{\sinh((2\beta+2\alpha)x/4)}{\sinh((\beta+\alpha)x/4)}=2\cosh((\beta+\alpha)x/4)
\end{eqnarray}

in the formula (\ref{D}) for $X_2(k,\alpha)$  from the very beginning and avoid the questions about continuity of the function. 

{\bf Remark 2, on tables.} The entries of the tables \ref{tab:x2acl} and \ref{tab:x2aex}  for a given algebra and $k$ are the  representation(s), denoted by highest weight, the quantum dimension of which is given by our main formula (\ref{D}). 

{\bf Remark 3, on the connection with the dimension formulae \cite{Cohen}}. In the $x\rightarrow 0 \,\,\,$ limit $X_2(x,k,\alpha,\beta,\gamma)$ gives the universal dimension formulae. When considered on the exceptional line by taking $\alpha=y,\beta=1-y,\gamma=2$ and for $k=2$, in the $x\rightarrow 0 \,\,\,$ limit the expression for $X_2(x,k,\alpha,\beta,\gamma)$ gives the following dimension formula 

\begin{eqnarray} \label{HH}
-\frac{10 (y-6) (y-5) (y+3) (y+4) (y+5) (2 y-5) (3 y-4) (5 y-6)}{(1-2 y)^2 (y-1)^3 y^4 (3 y-2)}
\end{eqnarray}

which coincides exactly with the universal formula on the exceptional line of \cite{Cohen} for representation $H$.

{\bf Remark 4, on $sp(2n)$ case.} We assume the following interpretation of this case. The point is that Vogel's parameters for $sp(2n)$ algebras can be obtained from those of $so(2n)$ by transformation $(\alpha,\beta,\gamma) \rightarrow (-1/2)(\beta,\alpha,-\gamma)$, which includes transposition of $\alpha$ and $\beta$. And indeed, we see in the table \ref{tab:x2bex}, that our formula gives quantum dimensions of some sequence of representations of $sp(2n)$, which however are not the Cartan powers of its $X_2$ representation. At the same time, table \ref{tab:x2bex} doesn't represent quantum dimensions of any new representations of $so$. We conclude, that the role of $X_2$ sequence of representations for $sp$ in our formulae is played by other series, given in the table \ref{tab:x2bex}.

\section{Permutations in the main formula}\label{sing}

The common feature of all universal formulae is that they should give a reasonable results when taken with permuted parameters. This is due to the fact that initially all universal parameters are on an equal footing."Reasonable" for e.g. dimension formulae means that they give dimensions for other representations of the same algebra. Sometimes they give dimensions of virtual representations, i.e. dimensions of true representations with minus sign. It happens in all known cases, and we show that for our case of quantum dimensions, it also does.  We present our formulae for all possible permutations of the universal parameters in a sequence of tables. The cases of virtual representations are also denoted by highest weights, but with minus sign. 

Our check mainly extends to the level of dimensions of representations, with some random checks on quantum dimensions' level.

The values of $X_2(k,\beta)$ for algebras on the exceptional line are presented in the table \ref{tab:x2bex}. Here we see a new phenomena: the value of $X_2(k,\beta)$ on, say point $k=2$ for $D_4$ algebra  (i.e. $\alpha=-2, \beta=4, \gamma=4$)   is not defined, since the limit of $0/0$  ambiguity depends on the way of approaching that point. However, if one approaches that point by one of the relevant lines, $Exc$ or $so$, the reasonable results are obtained. They are noted in the table and are in agreement with other cases, e.g. the result for $D_4$ agrees with the results for general orthogonal algebras in the table \ref{tab:x2acl}.

There are other cases, e.g. $k>3$ for $D_4$, when the double limit doesn't exist, however, the limit exists and is unique when restricted to any line, approaching the point $D_4$. We don't specially mention that cases in our tables.

When restricted to the exceptional line, with the identification of parameters as in Remark 3 above, the dimension formula, following from $X_2(2,\beta)$ in the $x \rightarrow 0$ limit coincides with that of $H^*$ of \cite{Cohen}, both are obtained from $H$ (\ref{HH}) by substitution $y \leftrightarrow 1-y$. 

We see that $X_2(3,\beta)$ for $D_4$ gives pure number $3$, independent on $x$. This should be interpreted as quantum dimension  of some representation of semidirect product of $D_4$ and its Dynkin diagram's automorphism group $S_3$. We assume that the corresponding representation is the trivial one for $D_4$ factor and the non-trivial reducible three-dimensional  permutation representation of $S_3$ factor. 
  
\FloatBarrier
   \begin{table} 
   \caption{$X_2(k,\beta)$ for the exceptional algebras}
       \begin{tabular}{|c|c|c|c|c|c|}
       \hline
            $k$&1&2&3&4&$\geq5$\\
            \hline
         $G_2$&$3\omega_1$&0&0&0&0\\
         \hline
         $F_4$&$\omega_2$&$3\omega_4$&0&0&0\\
         \hline
         $E_6$&$\omega_3$&$3\omega_1\oplus 3\omega_5$&$-(\omega_1+
         \omega_5)$&0&0\\
         \hline
         $E_7$&$\omega_2$&0&$-\omega_2$&-1&0\\
         \hline
      $E_8$&$\omega_6$&0&$\omega_8$&0&0\\
         \hline
         $D_4$& $\omega_1+\omega_3+\omega_4$& $\left\{\begin{tabular}{c}
           $\omega_1+\omega_3+\omega_4$ \, \text {on the} $Exc$ \text{ line}\\
           0 \, \text{on the}\, $so$ \, \text{line}
           \end{tabular} \right. $ &
         $\left\{\begin{tabular}{c}
                    3 \, \text {on the} $Exc$ \text{ line}\\
                    0 \, \text{on the}\, $so$ \, \text{line}
                    \end{tabular} \right. $&0&0\\
         \hline
         \end{tabular}
         \label{tab:x2bex}
      \end{table}   
  \FloatBarrier

  $X_2(k,\gamma)$ for the exceptional algebras are given in table \ref{tab:x2gex}.

  \FloatBarrier
 \begin{table}
 \caption{$X_2(k,\gamma)$ for the exceptional algebras}
     \begin{tabular}{|c|c|c|c|c|}
     \hline 
          $k$&1&2&3&$\geq4$\\
          \hline
       $G_2$&$3\omega_1$&$3\omega_1$&0&0\\
       \hline
       $F_4$&$\omega_2$&$\omega_2$&0&0\\
       \hline
       $E_6$&$\omega_3$&$\omega_3$&0&0\\
       \hline
       $E_7$&$\omega_2$&$\omega_2$&0&0\\
       \hline
    $E_8$&$\omega_6$&$\omega_6$&0&0\\
       \hline
       $D_4$& $\omega_1+\omega_3+\omega_4$& $\left\{\begin{tabular}{c}
           $\omega_1+\omega_3+\omega_4$ \, \text {on the} $Exc$ \text{ line}\\
           0 \, \text{on the}\, $so$ \, \text{line}
           \end{tabular} \right. $ &
         $\left\{\begin{tabular}{c}
                    0 \, \text {on the} $Exc$ \text{ line}\\
                    0 \, \text{on the}\, $so$ \, \text{line}\end{tabular}\right.$&0\\
                    \hline
       \end{tabular}
       \label{tab:x2gex}
 \end{table}
 \FloatBarrier

Again, when restricted to the exceptional line $\alpha=y,\beta=1-y,\gamma=2$ and in the limit $x \rightarrow 0$, 
 $X_2(2,\gamma)$ gives the following formula 

$$
\frac{5 (y-6) (y-4) (y+3) (y+5)}{(y-1)^2 y^2}
$$

which coincides with the dimension formula for $X_2$ from \cite{Cohen}, in agreement with the table \ref{tab:x2gex}.

    \FloatBarrier
 \begin{table}
 \caption{$X_2(k,\gamma)$ for the classical algebras}
     \begin{tabular}{|c|c|c|c|c|}
     \hline
          $k$&1&2&3&$\geq4$\\
          \hline
       $A_1$&0&$-2\omega$ on the $sl$ line&0&0\\
          \hline
       $A_2$&$3\omega_1\oplus 3\omega_2$&$-(\omega_1+\omega_2)$&0&0\\
          \hline
       $A_n,n\geq3$&$(2\omega_1+\omega_{n-1})\oplus
       (\omega_2+2\omega_{n})$&$-(\omega_1+\omega_n)$&0&0\\
          \hline
       $B_2$&$\omega_1+2\omega_2$&0 \text{on the $so$ line}&0 \text{on the $so$ line}&0\\
          \hline
       $B_3$&$\omega_1+2\omega_3$&0&0&0\\
          \hline
       $B_n,n\geq4$&$\omega_1+\omega_3$&0&0&0\\
          \hline
       $C_n,n\geq3$&$2\omega_1+\omega_2$&0&0&0\\
          \hline
       $D_4$&$\omega_1+\omega_3+\omega_4$&0 \text{on the $so$ line}&0 \text{on the $so$ line}&0\\
          \hline
       $D_5$&$\omega_1+\omega_3$&0&0&0\\
          \hline
       $D_6$&$\omega_1+\omega_3$&0 \text{on the $so$ line}&0&0\\
          \hline                
       $D_n,n\geq7$&$\omega_1+\omega_3$&0&0&0\\
          \hline
    \end{tabular}
    \label{tab:x2gcl}
\end{table}
    \FloatBarrier
    
    For the classical algebras $X_2(k,\beta)$ is given in the table \ref{tab:x2bcl}. For small ranked algebras there shows up a complicated picture, so we present the stabilized answers for sufficiently large ranks. The boundary depends on $k$, the larger $k$, the larger the boundary. At least the rank should be large enough to allow the existence of the fundamental weights mentioned in the table. 
    \FloatBarrier
 \begin{table}
 \caption{$X_2(k,\beta)$ for the classical algebras for sufficiently large $n$ (depends on $k$)}
     \begin{tabular}{|c|p{2.1cm}|p{2cm}|p{2cm}|p{2cm}|c|}
     \hline
          $k$&1&2&3&4&$\geq5$\\
          \hline
       $A_n$&$ (2\omega_1+\omega_{n-1}) \oplus (2\omega_{n}+\omega_{2})$ &$ (2\omega_2+\omega_{n-3})\oplus (2\omega_{n-1}+\omega_{4})$&$(2\omega_3+\omega_{n-5})\oplus (2\omega_{n-2}+\omega_{6})$&$(2\omega_4+\omega_{n-7})\oplus (2\omega_{n-3}+\omega_{8})$&$\cdots$\\
       \hline
       $B_n$&$\omega_1+\omega_3$&0  & $0$&$0$&$0$\\
       \hline
       $C_n$&$2\omega_1+\omega_2$&$2\omega_2+\omega_4$&$2\omega_3+\omega_6$&$2\omega_4+\omega_8$&$\cdots$\\
       \hline
       $D_n$&$\omega_1+\omega_3$&0&$0$&$0$&$0$\\
       \hline
    \end{tabular}
    \label{tab:x2bcl}
    \end{table}
  \FloatBarrier


\section{Conclusion}

The present results are of interest for the representation theory of simple Lie algebras. They hint on the interpretation of the universal formulae as result of some duality transformation between rank $N$ and Cartan power $k$. That transformation is based on a formula used in the Appendix, which  in its simplest form is an identity

\begin{eqnarray}
\prod_{i=1}^N \frac{k+i}{i}=
\prod_{i=1}^k \frac{N+i}{i}
\end{eqnarray}

The analogy of this transformation is that in \cite{MV11}, where for the proof of $N \leftrightarrow -N$ duality of $SO(2N)$ and $Sp(2N)$ one uses the so called {\it Maya} parametrization of Young diagrams, aimed to remove explicit rank $N$ from the range of the products over the running index. 

Simultaneously, the very existence of universal formulae for an arbitrary representation in the decomposition of powers of the adjoint is not proven. 

The present research is a part of a larger project, aimed to construct universal knot polynomials. In \cite{MMM} it is shown, that for 2- and 3-strand torus knots adjoint knot polynomials of different types can be presented in the universal form. This requires representation of quantum dimensions of all representations in a decomposition of square and cube of adjoint representation in the universal form. So, at $k=2$ our result gives a universal formula for one of the representations in the fourth power of the adjoint. Most of other representations also have a universal dimension formula. We assume that it is possible to obtain universal formulae for all other representations also, so the 4-th strand torus knots invariant polynomials will be possible to be represented in the universal form.

\section{Acknowledgments}

The work of MA was fulfilled within the Regional Doctoral Program on Theoretical and Experimental Particle Physics Program 
sponsored by VolkswagenStiftung.
The work of MA and RM is partially supported by the Science Committee of the Ministry of Science 
and Education of the Republic of Armenia under contract  18T-1C229.

\section{Appendix. Proof of the main formula}

\subsection{Notation} 
Below we omit the numerous $\sinh$ signs and 
 use the following notation instead:

\begin{eqnarray}
a\sinh\left[x: \right. \,\, \frac{A\cdot B...}{M\cdot N...}\equiv a\frac{\sinh(xA)\sinh(xB)...}{\sinh(xM)\sinh(xN)...}
\end{eqnarray}
where $x,a,A,B,...,M,N,...$ are numbers (dots between are not necessary, provided no ambiguity arises). For example

\begin{eqnarray}
2\sinh\left[\frac{x}{4}: \right. \,\, \frac{1\cdot 4}{2}\equiv 2\frac{\sinh(\frac{x}{4})\sinh(\frac{4x}{4})}{\sinh(\frac{2x}{4})}
\end{eqnarray}

One can derive simple rules which this notation obeys. E.g. 

\begin{eqnarray}
\left( \sinh\left[x: \right. \,\, A\cdot B \right) \left( \sinh \left[x: \right. M\cdot N\right) = \sinh\left[x: \right. \,\, A\cdot B \cdot M\cdot N
\end{eqnarray}

Of course, our notation belongs to the field of q-calculus, however we didn't find this or similar convenient notation, perhaps missed
 that.

Evidently, one gets the universal dimension formulae, just by omitting the front $\sinh$ sign for $L_?$-s and $X_2(x,k,\alpha,\beta,\gamma)$ in formulae below.

\subsection{$L_?$-s in the new notations}

The $L_?$ multipliers in the new notation take the following form  

$$
L_{11}=\sinh\left[\frac{x}{4}: \right. \prod _{i=1}^k \frac{
-\alpha (i-4)+2
\beta+\gamma}{\alpha i}
$$
$$L_{12}=\sinh\left[\frac{x}{4}: \right.\prod _{i=1}^k \frac{
\alpha (i-3)-2 \gamma}{ 
\alpha (i+1)}
$$

$$L_{13}=\sinh\left[\frac{x}{4}: \right.\prod _{i=1}^k \frac{
-\alpha
(i-2)+\beta+\gamma}{\beta-
\alpha
(i-2)}
$$
$$L_{14}=\sinh\left[\frac{x}{4}: \right.\prod _{i=1}^k\frac{-\alpha (i-3)+\beta+\gamma}
{ \alpha
(-i)+\alpha+\beta} 
$$
$$L_{15}=\sinh\left[\frac{x}{4}: \right.\prod _{i=1}^k\frac{  \left(\alpha
(i-3)-2
\beta\right)  
\left(\alpha (i-2)-2 \beta\right)
}{ \left(\gamma-\alpha
(i-2)\right)  \left(\alpha
(-i)+\alpha+\gamma\right)} $$

$$L_{21}=\sinh\left[\frac{x}{4}: \right.\prod _{i=1}^{2 k} \frac{  
   \alpha (i-5)-2
(\beta+\gamma)}{     
\alpha
(i-2)-2
\beta}
$$

$$L_{22}=\sinh\left[\frac{x}{4}: \right.\prod _{i=1}^{2 k} \frac{     
-\alpha
(i-3)+\beta+2 \gamma}{     
-\alpha
(i-2)+\beta+\gamma}$$

$$L_{23}=\sinh\left[\frac{x}{4}: \right.\frac{\alpha (3-2 k)+2
\beta+\gamma}{3 \alpha+2
\beta+\gamma} $$

$$L_{3}=\sinh\left[\frac{x}{4}: \right.\frac{ 
\left(  
3 \alpha (k-1)-2 (\beta+\gamma)\right)  
\left(  \alpha (3
k-4)-2
(\beta+\gamma) \right)}{ \left(4
\alpha+2\beta+2\gamma \right)
 \left( 
3 \alpha+2 (\beta+\gamma)  \right)} $$

$$L_{corr}=\sinh\left[\frac{x}{4}: \right.\prod _{i=1}^k\frac{\alpha
(-(i+k-4))+2
\beta+\gamma}{\alpha
(i+k-2)-2
\gamma}$$

The proof of the main
formula  is carried out case by case:  
for each set of parameters 
$\alpha, \beta, \gamma$ from Vogel's table  (except  $C_n$) we
compare the expression (\ref{D})  with the Weyl quantum
dimension (\ref{W}) for the corresponding algebra.

\subsection{$A_{N-1}$}
First we carry out the proof for the $A_{N-1}$ algebra.
\newline Substituting $\alpha=-2,\beta=2,\gamma=N$
in the 
$L$-terms, one gets
$$L_{11}=\sinh\left[\frac{x}{4}: \right.
\frac{  
\left(   N-2  \right)\cdot  
  N \dots  
\left(N+2k-4 \right)}{ 2\cdot4\dots 2k },$$
$$L_{12}=\sinh\left[\frac{x}{4}: \right.
\frac{  
\left(2N-4\right)\cdot  
\left(2N-2\right)\dots  
\left(2N+2k-6 \right)}{4\cdot6 \dots (2k+2)},$$
$$L_{13}=\sinh\left[\frac{x}{4}: \right.
\frac{  
  N \cdot  
\left(N+2 \right)\dots  
\left(N+2k-2  \right)}{(\alpha+\beta)\cdot2\dots (2k-2)},$$
$$L_{14}=\sinh\left[\frac{x}{4}: \right.
\frac{  
\left(N-2\right)\cdot  
  N \dots  
\left(N+2k-4\right)}{ 2\cdot 4 \dots 2k},$$
$$L_{15}=\cdot \sinh\left[\frac{x}{4}: \right.
\frac{ (2\alpha+2\beta)\cdot
   2^2\cdot 4^2 \dots 
(2k-2)^2 \cdot  
 2k }{ (N-2)
\cdot N^2 \dots
(N+2k-4)^2\cdot  \left(N+2k-2\right)},$$
$$L_{21}=\sinh\left[\frac{x}{4}: \right.
\frac{  
\left(2N-4\right)\cdot  
\left(2N-2\right)\dots  
\left(2N+2k-6 \right)\dots  
\left(2N+4k-6\right)}{ 2\cdot4\dots 2k\dots 4k},$$
$$L_{22}=\sinh\left[\frac{x}{4}: \right.
\frac{  \left( 2N-2 \right)\cdot   2N \dots  \left(  
2N+2k-4 \right)\dots  \left(  2N+4k-4  \right)}{  N
\cdot (N+2) \dots
(N+2k-2)\dots  (N+4k-2)},$$
$$L_{23}=\sinh\left[\frac{x}{4}: \right.\frac{ N+4k-2}
{N-2} $$
$$L_{3}=\sinh\left[\frac{x}{4}: \right.\frac{  \left(2N+6k-4 \right)\cdot
  \left(2N+6k-2\right)}
{  \left(2N-4 \right)\cdot
  \left(2N-2 \right)}$$
$$L_{corr}=\sinh\left[\frac{x}{4}: \right.
\frac{  
\left( N+2k-2\right)\cdot  
\left( N+2k\right)\dots  
\left(N+4k-4\right)}{  \left(2N+2k-2\right)\cdot  \left(2N+2k \right)
\dots  \left
(
2N+4k-4\right)}.$$

Notice, that
$$L_{13}\cdot L_{14}\cdot
L_{15}=2\cosh{\frac{\alpha+\beta}{4}}=2$$
Next we take the product of $L_{22},L_{corr},L_{23},L_3$: \newline

    \begin{multline*}
    L_{22}\cdot L_{corr}\cdot L_{23}\cdot
L_3=\\\sinh\left[\frac{x}{4}: \right.
\frac{  \left(  
2N-2  \right)\cdot   2N
 \dots  \left( 
2N+2k-4  \right)}{ N \cdot  \left(N+2\right)
\dots  \left(N+2k-4\right)\cdot  \left(N+4k-2
\right)}\cdot
\frac{N+4k-2}{ N-2}\cdot\\
\cdot
\frac{  \left(2N+6k-4  \right)\cdot
  \left(2N+6k-2\right)}
{  \left(2N-4 \right)\cdot
  \left(2N-2\right)}={}\\
 \sinh\left[\frac{x}{4}: \right.\frac{ 2N\cdot(2N+2)\dots(2N+2k-4)\cdot(2N+6k-4)(2N+6k-2)}
{(N-2) N \dots
(N+2k-4)(2N-4)}
\end{multline*}

Now we take the product of the remaining $L_{11},L_{12},
L_{21}$ terms.
{\scriptsize
\begin{multline*}
    L_{11}\cdot L_{12}\cdot L_{21}=\\\sinh\left[\frac{x}{4}: \right.
    \frac{(N-2)N\dots(N+2k-4)(2N-4)^2\dots(2N+2k-6)^2
    (2N+2k-4)(2N+2k-4)\dots(2N+4k-6)}{2\cdot4^2\cdot6^2\dots
    (2k)^2\cdot(2k+2)\cdot2\cdot4\cdot6\dots4k}.
\end{multline*}
}
and finally get
\begin{multline*}
   X_2(x,k,-2,2,N)=\\
   =L_{11}\cdot L_{12}\cdot L_{21}\cdot L_{22}\cdot L_{corr}\cdot L_{23}\cdot
L_3\cdot_{13}\cdot L_{14}\cdot L_{15}=\\
2\cdot \sinh\left[\frac{x}{4}: \right.
\frac{(2N-4)(2N-2)^2(2N)^3\dots(2N+2k-6)^3(2N+2k-4)^2}
 {2^2\cdot4^3\dots(2k)^3}\cdot\\
 \frac{(2N+2k-2)\dots(2N+4k-6)(2N+6k-4)(2N+6k-2)}
 {(2k+2)^2\cdot(2k+4)\dots4k}
\end{multline*}

Now we use the Weyl quantum dimension formula for the 
$k$-th Cartan power of the $X_2$ representation for the
$A_{N-1}$ algebra, which is the sum of the quantum dimensions of two irreps, constituing $X_2$ for $A$ series. Actually, quantum dimensions of these two representations are equal, since they are connected by automorphism of the Dynkin diagram.  \newline
Having 
$$D_Q=\prod _{\alpha>0} \frac{\sinh
\left(\frac{x}{2}((\lambda,\alpha)+(\lambda,\rho)\right)}
{\sinh\left(\frac{(\alpha,\rho)x}{2}\right)}$$
and taking into account, that the highest weight of 
one of representation is $\lambda=2k\omega_1+k\omega_{N-2}$, one
gets

\begin{multline*}
   D_Q^{A_{N-1}}=\prod _{i=1}^{N-2}
   \frac{\sinh
\left(\frac{x}{2}(k+i\right)}
{\sinh\left(\frac{ix}{2}\right)}\cdot
\prod _{i=2}^{N-3} \frac{\sinh
\left(\frac{x}{2}(k+i\right)}
{\sinh\left(\frac{ix}{2}\right)}\cdot \\
\prod _{i=1}^{N-3} \frac{\sinh
\left(\frac{x}{2}(2k+i\right)}
{\sinh\left(\frac{ix}{2}\right)}\cdot
\prod _{i=N-2}^{N-1} \frac{\sinh
\left(\frac{x}{2}(3k+i\right)}
{\sinh\left(\frac{ix}{2}\right)}
\end{multline*}
In this formula the product is taken over $N$, while the $X_2(x,k,-2,2,N)$ is represented
by several products over $k$.
We switch from one
product to another by the following identity:
\begin{equation}\label{Ntok}
    \prod_{i=1}^N \frac{\sinh
\left(\frac{x}{2}(k+i\right)}
{\sinh\left(\frac{ix}{2}\right)}=
\prod_{i=1}^k \frac{\sinh
\left(\frac{x}{2}(N+i\right)}
{\sinh\left(\frac{ix}{2}\right)}.
\end{equation}

In our notation we can write
\begin{multline*}
    D_Q^{A_{N-1}}=\sinh\left[\frac{x}{4}: \right.\prod_{i=1}^k \frac{2N-4+2i}{2i}\cdot
    \frac{2}{2k+2}\prod_{i=1}^k \frac{2N-6+2i}{2i}\cdot\\
    \prod_{i=1}^{2k} \frac{2N-6+2i}{2i}\cdot
    \frac{(6k+2N-4)(6k+2N-2)}{(2N-4)(2N-2)}
\end{multline*}
Then, after expanding the product signs and combining the 
multipliers, we get
\begin{multline*}
\sinh\left[\frac{x}{4}: \right. \frac{(2N-4)(2N-2)^2(2N)^3\dots(2N+2k-6)^3(2N+2k-4)^2}
 {2^2\cdot4^3\dots(2k)^3}\times\\
 \frac{(2N+2k-2)\dots(2N+4k-6)(2N+6k-4)(2N+6k-2)}
 {(2k+2)^2\cdot(2k+4)\dots4k}
\end{multline*}
which is equal to the expression  $1/2\cdot X_2(x,k,-2,2,N)$, as
expected. So, we have proven the formula (\ref{D}) for $A_{N-1}$
algebra.

\subsection{$B_N$}

 For this case we should substitute $\alpha=-2,\beta=4,\gamma=2N-3$, so 
 
$$L_{11}=\sinh\left[\frac{x}{4}: \right.
\frac{(2N-1)(2N+1)\dots\ (2N+2k-3)}{2
\cdot4\dots2k},$$
$$L_{12}=\sinh\left[\frac{x}{4}: \right.\frac{(4N-10)(4N-8)\dots\ (4N+2k-12)}{4
\cdot6\dots(2k+2)}
,$$
$$L_{13}=\sinh\left[\frac{x}{4}: \right.\frac{(2N-1)(2N+1)\dots\ (2N+2k-3)}{2
\cdot4\dots2k}
,$$
$$L_{14}=\sinh\left[\frac{x}{4}: \right.\frac{(2N-3)(2N-1)\dots\ (2N+2k-5)}{4
\cdot6\dots(2k+2)}
,$$
$$L_{15}=\sinh\left[\frac{x}{4}: \right.\frac{4
\cdot6^2\cdot8^2\dots(2k+2)^2\cdot(2k+4)}
{(2N-5)(2N-3)^2\dots\ (2N+2k-7)^2(2N+2k-5)}
,$$
$$L_{21}=\sinh\left[\frac{x}{4}: \right.\frac{(4N-6)(4N-4)\dots\ (4N+4k-8)}{6
\cdot8\dots(4k+2)}
,$$
$$L_{22}=\sinh\left[\frac{x}{4}: \right.\frac{(4N-6)(4N-4)\dots\
(4N+4k-8)}{(2N-1)(2N+1)\dots\ (2N+4k-3)}
,$$
$$L_{23}=\sinh\left[\frac{x}{4}: \right.\frac{2N+4k-1}{2N-1}$$
$$L_{3}=\sinh\left[\frac{x}{4}: \right.\frac{(4N+6k-4)(4N+6k-6)}{(4N-6)(4N-4)}$$
$$L_{corr}=\sinh\left[\frac{x}{4}: \right.\frac{(2N+2k-1)(2N+2k+1)\dots\
(2N+4k-3)}{(4N+2k-8)(4N+2k-6)\dots\ (4N+4k-10)}.$$

The product of $L_{11},L_{14},L_{15}$ gives

$$
  L_{11}\cdot L_{14}\cdot L_{15} =\sinh\left[\frac{x}{4}: \right.
  \frac{(2k+2)(2k+4)(2N+2k-5)(2N+2k-3)}{2\cdot4\cdot
  (2N-5)(2N-3)}.$$
The product of $L_{22}, L_{corr}, L_{13}$ gives
$$L_{22}\cdot L_{corr} \cdot L_{13}=\textit{sinh}
\left[ \right.\frac{(4N-6)(4N-4)\dots\
(4N+2k-8)(4N+4k-8)}{2\cdot 4\dots\ 2k}$$
And the product of remaining four terms is
{\scriptsize
\begin{multline*}
L_{12}\cdot L_{21}\cdot L_{23}\cdot L_3
=\\\sinh\left[\frac{x}{4}: \right.
\frac{(4N-10)(4N-8)(4N-6)^2(4N-4)^2(4N-2)^3
\dots (4N+2k-12)^3(4N+2k-10)^2(4N+2k-8)^2}{(2N-5)(2N-3)(2N-1)\cdot2^2}
\times\\
\frac{(4N+2k-6)\dots(4N+4k-8)(2N+4k-1)(2N+2k-5)(2N
+2k-3)
(4N+6k-6)(4N+6k-4)}{4^3
\dots(2k)^3\cdot(2k+2)\dots(4k+4)}
\end{multline*}
}

So, the product of all  $L$-terms is:
{\scriptsize
\begin{multline}\label{x2B}
     X_2(x,k,-2,4,2N-3)=
  L_{11}\cdot L_{12}\cdot L_{21}\cdot
  L_{22}\cdot L_{corr} \cdot
  L_{23}\cdot L_3\cdot L_{13}\cdot L_{14}\cdot
  L_{15}=\\\sinh\left[\frac{x}{4}: \right.
\frac{(4N-10)(4N-8)(4N-6)^2(4N-4)^2(4N-2)^3
\dots (4N+2k-12)^3(4N+2k-10)^2(4N+2k-8)^2}{(2N-5)(2N-3)(2N-1)\cdot2^2}
\times\\
\frac{(4N+2k-6)\dots(4N+4k-8)(2N+4k-1)(2N+2k-5)(2N
+2k-3)
(4N+6k-6)(4N+6k-4)}{4^3
\dots(2k)^3\cdot(2k+2)\dots(4k+4)}. 
\end{multline}
}

Now we turn to the Weyl
formula for $B_N$ algebra:

\begin{multline*}
    D_Q^{B_N}=\sinh\left[\frac{x}{4}: \right.\prod_{i=1}^{N-2} \frac{2k+2i}{2i}\cdot
    \frac{2k+2N-3}{2N+3}\prod_{i=N}^{2N-4} \frac{2k+2i-2}{2i-2}\cdot\\
    \prod_{i=1}^{N-3} \frac{2k+2i}{2i}\cdot
    \frac{2N+2k-5}{2N-5}\cdot \prod_{i=N-1}^
    {2N-5}\frac{2k+2i-2}{2i-2}\cdot
    \frac{2k+4}{4}\cdot\\
    \prod_{i=3}^{N-1} \frac{4k+2i}{2i}\cdot
    \frac{4k+2N-1}{2N-1}\cdot
    \prod_{i=N+1}^{2N-3}
    \frac{4k+2i-2}{2i-2}\cdot\\
    \frac{(4k+4N-8)(6k+4N-6)(6k+4N-4)}
    {(4N-8)(4N-6)
    (4N-4)}.
\end{multline*}
With the help of (\ref{Ntok}) identity  we switch
the products limits to  $k$:

 $$ \prod_{i=1}^{N-2} \frac{2k+2i}{2i}=
   \prod_{i=1}^{k} \frac{2i+2N-4}{2i}=
   \frac{(2N-2)2N\dots (2N+2k-4)}{2
   \cdot 4 \dots2k},    
   $$
 $$ \prod_{i=1}^{N-3} \frac{2k+2i}{2i}=
   \prod_{i=1}^{k} \frac{2i+2N-6}{2i}=
   \frac{(2N-4)(2N-2)\dots (2N+2k-6)}{2
   \cdot 4 \dots2k},    
   $$
  
  \begin{multline*}
  \prod_{i=3}^{N-1} \frac{4k+2i}{2i}=
   \prod_{i=1}^{2k} \frac{2i+2N-2}{2i}\cdot 
   \frac{2\cdot 4}{(4k+2)(4k+4)}= \\
   \frac{2\cdot 4}{(4k+2)(4k+4)}\cdot 
   \frac{2N(2N+2)\dots (2N+4k-2)}{2
   \cdot 4 \dots4k},    
   \end{multline*} 
   
$$ \prod_{i=N}^{2N-4} \frac{2k+2i-2}{2i-2}=
   \prod_{i=N-1}^{2N-5} \frac{2k+2i}{2i}=
   \prod_{i=1}^{k} \frac{4N-10+2i}{2i}\cdot
   \prod_{i=1}^{N-2} \frac{2i}{2i+2k}=$$
   
  $$ =\prod_{i=1}^{k} \frac{4N-10+2i}{2i}\cdot
   \prod_{i=1}^{k} \frac{2i}{2i+2N-4}=
   \frac{(4N-8)(4N-6)\dots (4N+2k-10)}{(2N-2)
   \cdot 2N \dots(2N+2k-4)},    
   $$
   
$$ \prod_{i=N-1}^{2N-5} \frac{2k+2i-2}{2i-2}=
   \prod_{i=N-2}^{2N-6} \frac{2k+2i}{2i}=
   \prod_{i=1}^{k} \frac{4N-12+2i}{2i}\cdot
   \prod_{i=1}^{N-3} \frac{2i}{2i+2k}=$$
   
  $$ =\prod_{i=1}^{k} \frac{4N-12+2i}{2i}\cdot
   \prod_{i=1}^{k} \frac{2i}{2i+2N-6}=
   \frac{(4N-10)(4N-8)\dots (4N+2k-12)}{(2N-4)
   \cdot (2N-2) \dots(2N+2k-6)},    
   $$
And the last term needed to be transformed:

$$ \prod_{i=N+1}^{2N-3} \frac{4k+2i-2}{2i-2}=
   \prod_{i=N}^{2N-4} \frac{4k+2i}{2i}=
   \prod_{i=1}^{2k} \frac{4N-8+2i}{2i}\cdot
   \prod_{i=1}^{N-1} \frac{2i}{2i+4k}=$$
   
  $$ =\prod_{i=1}^{2k} \frac{4N-8+2i}{2i}\cdot
   \prod_{i=1}^{2k} \frac{2i}{2i+2N-2}=
   \frac{(4N-6)(4N-4)\dots (4N+4k-8)}{2N
   \cdot (2N+2) \dots(2N+4k-2)}.  
   $$

Next, taking into account the transformations above, we
rewrite the expression for $D_{Q}^{B_N}$:

{\scriptsize
\begin{multline*}
    D_Q^{B_N}=\sinh\left[\frac{x}{4}: \right.\frac{(2N-2)2N\dots (2N+2k-4)}{2
   \cdot 4 \dots2k}\cdot
    \frac{2k+2N-3}{2N+3}\frac{(4N-8)(4N-6)\dots (4N+2k-10)}{(2N-2)
   \cdot 2N \dots(2N+2k-4)}\cdot\\
    \frac{(2N-4)(2N-2)\dots (2N+2k-6)}{2
   \cdot 4 \dots2k}\cdot
    \frac{2N+2k-5}{2N-5}\cdot \frac{(4N-10)(4N-8)\dots (4N+2k-12)}{(2N-4)
   \cdot (2N-2) \dots(2N+2k-6)}\cdot
    \frac{2k+4}{4}\cdot\\
   \frac{(4N-8)(4N-6)\dots (4N+2k-10)}{(2N-2)
   \cdot 2N \dots(2N+2k-4)}\cdot
    \frac{4k+2N-1}{2N-1}\cdot
    \frac{(4N-6)(4N-4)\dots (4N+4k-8)}
    {2N
   \cdot (2N+2) \dots(2N+4k-2)}
    \cdot\\
    \frac{(4k+4N-8)(6k+4N-6)(6k+4N-4)}
    {(4N-8)(4N-6)
    (4N-4)}.
  \end{multline*}
  }

Combining similar terms, one gets
{ \scriptsize 
\begin{multline*}
     D_Q^{B_N}=\\\sinh\left[\frac{x}{4}: \right.
     \frac{(4N-10)(4N-8)(4N-6)^2(4N-4)^2(4N-2)^3
\dots (4N+2k-12)^3(4N+2k-10)^2(4N+2k-8)^2}
     {(2N-5)(2N-3)(2N-1)\cdot2^2}\times\\
     \frac{
(4N+2k-6)\dots(4N+4k-8)(2N+4k-1)(2N+2k-5)(2N
+2k-3)
(4N+6k-6)(4N+6k-4) }{4^3
\dots(2k)^3\cdot(2k+2)\dots(4k+4)}
\end{multline*}
}
which coincides with $X_2(x,k,-2,4,2N-3)$ (\ref{x2B}).

\subsection{$C_N$}
It was mentioned above that for $k$'s greater than 1 our universal formula gives 0 values for $C_N$ algebra.
So now we observe the $k=1$ case. Substituting $\alpha=-2, \beta=1, \gamma=N+2$ in the $L_?$ terms, one has

$$L_{11}=\sinh\left[\frac{x}{4}: \right.
\frac{N-2}{2},$$
$$L_{12}=\sinh\left[\frac{x}{4}: \right.\frac{2N}{4}
,$$
$$L_{13}=\sinh\left[\frac{x}{4}: \right.\frac{N+1}{1}
,$$
$$L_{14}=\sinh\left[\frac{x}{4}: \right.\frac{N-1}{1}
,$$
$$L_{15}=\sinh\left[\frac{x}{4}: \right.\frac{(-2\beta-\alpha)
\cdot 2}
{N(N+2)}
,$$
$$L_{21}=\sinh\left[\frac{x}{4}: \right.\frac{(2N-2)2N}{(-2\beta-\alpha)\cdot 2}
,$$
$$L_{22}=\sinh\left[\frac{x}{4}: \right.\frac{(2N+1)(2N+3)}{(N+1)(N+3)}
,$$
$$L_{23}=\sinh\left[\frac{x}{4}: \right.\frac{N+2}{N-2}$$
$$L_{3}=\sinh\left[\frac{x}{4}: \right.\frac{(2N+6k-2)(2N+6k)}{(2N-2)2N}$$
$$L_{corr}=\sinh\left[\frac{x}{4}: \right.\frac{N}{2N+4}.$$

Then we take the product of all these terms

     \begin{multline*}
     X_2(x,k,-2,1,N+2) = L_{11}\cdot
     L_{12}\cdot L_{13}\cdot L_{14}\cdot
     L_{15}\cdot L_{21}\cdot L_{22}\cdot 
     L_{23}\cdot L_{3}\cdot L_{corr}=\\
         \sinh\left[\frac{x}{4}: \right.
         \frac{2N(N-1)(2N+1)(2N+3)(2N+6)}
     {1^2\cdot2\cdot4\cdot (N+3)},
     \end{multline*}    
     
which coincides with the expression one gets after calculating the quantum dimension of the $2\omega_1+\omega_2$ representation
 of $C_N$ algebra using the Weyl formula.

\subsection{$D_N$}
 For this case
we substitute $\alpha=-2,\beta=4,\gamma=2N-4$, so $L$-terms become
$$L_{11}=\sinh\left[\frac{x}{4}: \right.
\frac{(2N-2)2N\dots\ (2N+2k-4)}{2
\cdot4\dots2k},$$
$$L_{12}=\sinh\left[\frac{x}{4}: \right.\frac{(2N-6)(2N-5)\dots\ (2N+k-7)}{2
\cdot3\dots(k+1)}
,$$
$$L_{13}=\sinh\left[\frac{x}{4}: \right.\frac{(N-1)N\dots\ (N+k-2)}{1
\cdot2\dots k}
,$$
$$L_{14}=\sinh\left[\frac{x}{4}: \right.\frac{(N-2)(N-1)\dots\ (N+k-3)}{2
\cdot3\dots(k+1)}
,$$
$$L_{15}=\sinh\left[\frac{x}{4}: \right.\frac{2
\cdot3^2\cdot4^2\dots(k+1)^2\cdot(k+2)}
{(N-3)(N-2)^2\dots\ (N+k-4)^2(N+k-3)}
,$$
$$L_{21}=\sinh\left[\frac{x}{4}: \right.\frac{(2N-4)(2N-3)\dots\ (2N+2k-5)}{3
\cdot4\dots(2k+2)}
,$$
$$L_{22}=\sinh\left[\frac{x}{4}: \right.\frac{(2N-4)(2N-3)\dots\
(2N+2k-5)}{(N-1)N\dots\ (N+2k-2)}
,$$
$$L_{23}=\sinh\left[\frac{x}{4}: \right.\frac{N+2k-1}{N-1}$$
$$L_{3}=\sinh\left[\frac{x}{4}: \right.\frac{(2N+3k-4)(2N+3k-3)}{(2N-4)(2N-3)}$$
$$L_{corr}=\sinh\left[\frac{x}{4}: \right.\frac{(N+k-1)(N+k)\dots\
(N+2k-2)}{(2N+k-5)(2N+k-4)\dots\ (2N+2k-6)}.$$
The product of $L_{13},L_{14},L_{15}$ terms gives
$$
  L_{13}\cdot L_{14}\cdot L_{15} =\sinh\left[\frac{x}{4}: \right.
  \frac{(k+1)(k+2)(N+k-3)(N+k-2)}{1\cdot2\cdot
  (N-3)(N-2)}.$$
Then, 
{\scriptsize
$$L_{22}\cdot L_{corr} \cdot L_{11}\cdot L_3=\sinh\left[\frac{x}{4}: \right.
\frac{(2N-2)(2N-1)\dots\
(2N+k-6)\cdot(2N+2k-5)(2N+3k-4)(2N+3k-3)}
{1\cdot 2\dots\ k}$$
}

And the remaining ones
{\scriptsize
\begin{multline*}
   L_{12}\cdot L_{21}\cdot L_{23}
=\\
\sinh\left[\frac{x}{4}: \right.
\frac{(2N-10)(2N-5)(2N-4)^2
\dots (2N+k-7)^2(2N+k-6)\dots(2N+2k-5)(N+2k-1)}
{2\cdot3^2\dots(k+1)^2\cdot(k+2)\dots(2k+2)(N-1)}. 
\end{multline*}
}

Overall, for $X_2(x,k,-2,4,2N-4)$ one gets

{\scriptsize
\begin{multline}\label{x2D}
   X_2(x,k,-2,4,2N-4)=\\\sinh\left[\frac{x}{4}: \right.
  \frac{(N+k-3)(N+k-2)(2N-6)(2N-5)(2N-4)^2
(2N-3)^2(2N-2)^3
\dots (2N+k-7)^3}
     {(N-3)(N-2)(N-1)\cdot1^2\cdot2^3
\dots k^3\cdot(k+1)}\times\\
     \frac{(2N+k-6)^2(2N+k-5)\dots
(2N+2k-6)(2N+2k-5)^2(N+2k-1)(2N+3k-4)(N+3k-3) }{(k+3)\dots(2k+2)}.
\end{multline}
}

The Weyl formula for $D_N$ algebra is

{\scriptsize
\begin{multline*}
    D_Q^{D_N}=\sinh\left[\frac{x}{4}: \right.\prod_{i=1}^{2N-6} \frac{k+i}{i}\cdot
   \prod_{i=1}^{2N-7} \frac{k+i}{i}\cdot
   \frac{(k+2)(k+N-3)(k+N-2)}
    {2\cdot (N-3)
    (N-2)}\cdot
    \prod_{i=3}^{2N-5} \frac{2k+i}{i}\cdot\\
    \frac{(N+2k-1)(2N-5+2k)}{(N-1)(2N-5)}\cdot 
    \frac{(2N+3k-4)(2N+3k-3)}{(2N-4)(2N-3)}=\\
  \sinh\left[\frac{x}{4}: \right.  \prod_{i=1}^{k} \frac{2N-6+i}{i}\cdot
   \prod_{i=1}^{k} \frac{2N-7+i}{i}\cdot
   \frac{(k+2)(k+N-3)(k+N-2)}
    {2\cdot (N-3)
    (N-2)}\cdot\\
   \frac{1\cdot 2}{(2k+1)(2k+2)} 
   \prod_{i=1}^{2k} \frac{2N-5+i}{i}\cdot\\
    \frac{(N+2k-1)(2N-5+2k)}{(N-1)(2N-5)}\cdot 
    \frac{(2N+3k-4)(2N+3k-3)}{(2N-4)(2N-3)}=\\\sinh\left[\frac{x}{4}: \right.
  \frac{(N+k-3)(N+k-2)(2N-6)(2N-5)(2N-4)^2
(2N-3)^2(2N-2)^3
\dots (2N+k-7)^3}
     {(N-3)(N-2)(N-1)\cdot1^2\cdot2^3
\dots k^3\cdot(k+1)}\times\\
     \frac{(2N+k-6)^2(2N+k-5)\dots
(2N+2k-6)(2N+2k-5)^2(N+2k-1)(2N+3k-4)(N+3k-3) }{(k+3)\dots(2k+2)},
\end{multline*}
}

which coincides with  $X_2(x,k,-2,4,2N-4)$ (\ref{x2D}).

\subsection{$G_2$}
For $G_2$ exceptional algebra Vogel's parameters 
take  values $\alpha=-2, \beta=10/3, \gamma=8/3$.
Substituting them in the $L$-terms, one has

$$L_{11}^{G_2}=\sinh\left[\frac{x}{6}: \right.
\frac{5\cdot 8\dots\ (3k+2)}{3
\cdot6\dots3k},$$
$$L_{12}^{G_2}=\sinh\left[\frac{x}{6}: \right.\frac{2\cdot 5\dots\ (3k-1)}{6
\cdot9\dots(3k+3)}
,$$
$$L_{13}^{G_2}=\sinh\left[\frac{x}{6}: \right.\frac{6\cdot 9\dots\ (3k+3)}{2
\cdot5\dots (3k-1)}
,$$
$$L_{14}^{G_2}=\sinh\left[\frac{x}{6}: \right.\frac{3\cdot 6\dots\ 3k}{5
\cdot8\dots(3k+2)}
,$$
$$L_{15}^{G_2}=\sinh\left[\frac{x}{6}: \right.\frac{7
\cdot10\dots(3k+4)}
{1\cdot 4\dots\ (3k-2)}
,$$
$$L_{21}^{G_2}=\sinh\left[\frac{x}{6}: \right.\frac{6\cdot 9\dots\ (6k+3)}{7
\cdot10\dots(6k+4)}
,$$
$$L_{22}^{G_2}=\sinh\left[\frac{x}{6}: \right.\frac{7\cdot 10\dots\
(6k+4)}{6\cdot9\dots\ (6k+3)}
,$$
$$L_{23}^{G_2}=\sinh\left[\frac{x}{6}: \right.\frac{6k+5}{5}$$
$$L_{3}^{G_2}=\sinh\left[\frac{x}{6}: \right.\frac{(9k+9)(9k+6)}{6\cdot 9}$$
$$L_{corr}^{G_2}=1.$$
First  notice, that
$$L_{11}^{G_2}\cdot L_{14}^{G_2}=L_{12}^{G_2}\cdot L_{13}^{G_2}=
L_{21}^{G_2}\cdot L_{22}^{G_2}=1$$
Thus, in the expression for $X_2(x,k,-2,10/3,8/3)$ only the 
$L_{15}^{G_2}, L_{23}^{G_2},
L_{3}^{G_2}$ terms contribute, so that

\begin{multline*}
 X_2(x,k,-2,10/3,8/3)= L_{15}^{G_2}\cdot L_{23}^{G_2}\cdot
L_{3}^{G_2} =\\=\sinh\left[\frac{x}{6}: \right.
  \frac{7
\cdot10\dots(3k+4)}
{1\cdot 4\dots (3k-2)}\cdot \frac{6k+5}{5}\cdot
\frac{(9k+9)(9k+6)}{6\cdot 9}=\\=\sinh\left[\frac{x}{4}: \right.
\frac{(3k+1)(3k+4)(6k+5)(9k+9)(9k+6)}{1\cdot 4\cdot 5\cdot 6 \cdot 9},
\end{multline*}
which coincides with the expression the Weyl 
formula (\ref{W}) gives for quantum dimension of $G_2$ algebra.

\subsection{$F_4$}
In this case we have $\alpha=-2, \beta=5, \gamma=6$
$$L_{11}^{F_4}=\sinh\left[\frac{x}{4}: \right.
\frac{10\cdot 12\dots\ (2k+8)}{2
\cdot4\dots2k}=\sinh\left[\frac{x}{4}: \right.\frac{(2k+2)(2k+4)(2k+6)(2k+8)}{2\cdot4\cdot6\cdot8},$$
$$L_{12}^{F_4}=\sinh\left[\frac{x}{4}: \right.
\frac{8\cdot 10\dots\ (2k+6)}{2
\cdot4\dots(2k+2)}=\sinh\left[\frac{x}{4}: \right.\frac{(2k+4)(2k+6)}{4\cdot6},$$
$$L_{13}^{F_4}=\sinh\left[\frac{x}{4}: \right.
\frac{9\cdot 11\dots\ (2k+7)}{3
\cdot5\dots(2k+1)}=\sinh\left[\frac{x}{4}: \right.
\frac{(2k+3)(2k+5)(2k+7)}{3\cdot5\cdot7},$$
$$L_{14}^{F_4}=\sinh\left[\frac{x}{4}: \right.
\frac{7\cdot 9\dots\ (2k+6)}{5
\cdot7\dots(2k+3)}=\sinh\left[\frac{x}{4}: \right.\frac{2k+5}{5},$$
$$L_{15}^{F_4}=\sinh\left[\frac{x}{4}: \right.
\frac{8\cdot 10\dots\ (2k+6)}{4
\cdot6\dots(2k+2)}=\sinh\left[\frac{x}{4}: \right.\frac{(2k+4)(2k+6)}{4\cdot6},$$
$$L_{21}^{F_4}=\sinh\left[\frac{x}{4}: \right.
\frac{14\cdot 16\dots\ (4k+12)}{8
\cdot10\dots(4k+6)}=\sinh\left[\frac{x}{4}: \right.
\frac{(4k+8)(4k+10)(4k+12)}{8\cdot10\cdot12},$$
$$L_{22}^{F_4}=\sinh\left[\frac{x}{4}: \right.
\frac{13\cdot 15\dots\ (4k+11)}{9
\cdot11\dots(4k+7)}=\sinh\left[\frac{x}{4}: \right.
\frac{(4k+9)(4k+11)}{9\cdot11},$$
$$L_{23}^{F_4}=\sinh\left[\frac{x}{4}: \right.\frac{4k+10}{10}$$
$$L_{3}^{F_4}=\sinh\left[\frac{x}{4}: \right.
\frac{(6k+14)(6k+16)}{14\cdot16},$$
$$L_{corr}^{F_4}=1.$$
The product of all these terms
$$X_2(x,k,-2,5,6) = L_{11}^{F_4}\cdot
     L_{12}^{F_4}\cdot L_{13}^{F_4}\cdot L_{14}^{F_4}\cdot
     L_{15}^{F_4}\cdot L_{21}^{F_4}\cdot L_{22}^{F_4}\cdot 
     L_{23}^{F_4}\cdot L_{3}^{F_4}\cdot L_{corr}^{F_4}=$$
     \begin{multline*}
         \sinh\left[\frac{x}{4}: \right.
         \frac{(2k+2)(2k+4)^3(2k+5)^2(2k+6)^3(2k+7)
     (2k+8)(4k+8)(4k+10)^2}
     {2\cdot3\cdot4^3\cdot 5^2\cdot 6^3\cdot 7\cdot 8^2}\times\\
    \times \frac{(4k+11)(4k+12)(6k+14)(6k+16)}{9\cdot 10^2 \cdot11\cdot12\cdot14\cdot16}
     \end{multline*}    
immediately coincides with expression Weyl formula gives for $F_4$ 
 algebra.
 
 \subsection{$E_6$}
 For $E_6$ the Vogel parameters are $\alpha=-2, \beta=6, \gamma=8$.
$$L_{11}^{E_6}=\sinh\left[\frac{x}{4}: \right.
\frac{7\cdot 8\dots\ (k+6)}{1
\cdot1\dots k}=\sinh\left[\frac{x}{4}: \right.
\frac{(k+1)(k+2)\dots(k+6)}{1\cdot2\dots6},$$
$$L_{12}^{E_6}=\sinh\left[\frac{x}{4}: \right.
\frac{6\cdot 7\dots\ (k+5)}{2
\cdot3\dots(k+8)}=\sinh\left[\frac{x}{4}: \right.
\frac{(k+2)(k+3)(k+4)(k+5)}{2\cdot3\cdot4\cdot5},$$
$$L_{13}^{E_6}=\sinh\left[\frac{x}{4}: \right.
\frac{6\cdot 7\dots\ (k+5)}{2
\cdot3\dots(k+8)}=\sinh\left[\frac{x}{4}: \right.
\frac{(k+2)(k+3)(k+4)(k+5)}{2\cdot3\cdot4\cdot5}
,$$
$$L_{14}^{E_6}=\sinh\left[\frac{x}{4}: \right.
\frac{5\cdot 6\dots\ (k+4)}{3
\cdot4\dots(k+2)}=\sinh\left[\frac{x}{4}: \right.
\frac{(k+3)(k+4)}{3\cdot4},$$
$$L_{15}^{E_6}=\sinh\left[\frac{x}{4}: \right.\frac{5\cdot 6\dots\ (k+4)}{3
\cdot4\dots(k+2)}=\sinh\left[\frac{x}{4}: \right.\frac{(k+3)(k+4)}{3\cdot4}
,$$
$$L_{21}^{E_6}=\sinh\left[\frac{x}{4}: \right.
\frac{10\cdot 11\dots\ (2k+9)}{5
\cdot6\dots(2k+4)}=\sinh\left[\frac{x}{4}: \right.
\frac{(2k+5)(2k+6)\dots(2k+9)}{5\cdot6\dots9},$$
$$L_{22}^{E_6}=\sinh\left[\frac{x}{4}: \right.
\frac{9\cdot 10\dots\ (2k+8)}{6
\cdot7\dots(2k+5)}=\sinh\left[\frac{x}{4}: \right.
\frac{(2k+6)(2k+7)(2k+8)}{6\cdot7\cdot8},$$
$$L_{23}^{E_6}=\sinh\left[\frac{x}{4}: \right.\frac{2k+7}{7}$$
$$L_{3}^{E_6}=\sinh\left[\frac{x}{4}: \right.\frac{(3k+10)(3k+11)}{10\cdot11},$$
$$L_{corr}^{E_6}=1.$$
The product of all these terms gives

$$X_2(x,k,-2,6,8) = L_{11}^{E_6}\cdot
     L_{12}^{E_6}\cdot L_{13}^{E_6}\cdot L_{14}^{E_6}\cdot
     L_{15}^{E_6}\cdot L_{21}^{E_6}\cdot L_{22}^{E_6}\cdot 
     L_{23}^{E_6}\cdot L_{3}^{E_6}\cdot L_{corr}^{E_6}=
     $$
 {\scriptsize
 $$ =\sinh\left[\frac{x}{4}: \right.\frac{(k+1)(k+2)^3(k+3)^5(k+4)^5(k+5)^3
     (k+6)(2k+5)(2k+6)^2(2k+7)^3(2k+8)^2(2k+9)}
     {1\cdot2^3\cdot3^5\cdot4^5\cdot 5^4\cdot 6^3\cdot 7^3\cdot 8^2\cdot
     9\cdot 10 \cdot11}$$
     }
     which coincides with quantum dimension $D_Q^{E_6}$ (\ref{W}).
     
     \subsection{$E_7$}
 For $E_7$ Vogel's parameters are $\alpha=-2, \beta=8, \gamma=12$.
$$L_{11}^{E_7}=\sinh\left[\frac{x}{4}: \right.
\frac{11\cdot 12\dots\ (k+10)}{1
\cdot1\dots k}=\sinh\left[\frac{x}{4}: \right.\frac{(k+1)(k+2)\dots(k+10)}{1\cdot2\dots10},$$
$$L_{12}^{E_7}=\sinh\left[\frac{x}{4}: \right.
\frac{10\cdot 11\dots\ (k+9)}{2
\cdot3\dots(k+1)}=\sinh\left[\frac{x}{4}: \right.
\frac{(k+2)(k+3)(k+4)\dots(k+9)}{2\cdot3\dots9},$$
$$L_{13}^{E_7}=\sinh\left[\frac{x}{4}: \right.
\frac{9\cdot 10\dots\ (k+8)}{3
\cdot4\dots(k+2)}=\sinh\left[\frac{x}{4}: \right.
\frac{(k+3)(k+4)(k+5)(k+6)(k+7)(k+8)}{3\cdot4\dots8}
,$$
$$L_{14}^{E_7}=\sinh\left[\frac{x}{4}: \right.
\frac{8\cdot 9\dots\ (k+7)}{4
\cdot5\dots(k+3)}=\sinh\left[\frac{x}{4}: \right.\frac{(k+4)(k+5)(k+6)(k+7)}{4\cdot5\cdot6\cdot7},$$
$$L_{15}^{E_7}=\sinh\left[\frac{x}{4}: \right.\frac{7\cdot 8\dots\ (k+6)}{5
\cdot6\dots(k+4)}=\sinh\left[\frac{x}{4}: \right.\frac{(k+5)(k+6)}{5\cdot6}
,$$
$$L_{21}^{E_7}=\sinh\left[\frac{x}{4}: \right.
\frac{16\cdot 17\dots\ (2k+15)}{7
\cdot8\dots(2k+6)}=\sinh\left[\frac{x}{4}: \right.
\frac{(2k+7)(2k+8)\dots(2k+15)}{7\cdot8\dots15},$$
$$L_{22}^{E_7}=\sinh\left[\frac{x}{4}: \right.
\frac{14\cdot 15\dots\ (2k+13)}{9
\cdot10\dots(2k+8)}=\sinh\left[\frac{x}{4}: \right.
\frac{(2k+9)\dots(2k+13)}{9\dots13},$$
$$L_{23}^{E_7}=\sinh\left[\frac{x}{4}: \right.\frac{2k+11}{11}$$
$$L_{3}^{E_7}=\sinh\left[\frac{x}{4}: \right.\frac{(3k+16)(3k+17)}{16\cdot17},$$
$$L_{corr}^{E_7}=1.$$
The product of all these terms gives

$$X_2(x,k,-2,8,12) = L_{11}^{E_7}\cdot
     L_{12}^{E_7}\cdot L_{13}^{E_7}\cdot L_{14}^{E_7}\cdot
     L_{15}^{E_7}\cdot L_{21}^{E_7}\cdot L_{22}^{E_7}\cdot 
     L_{23}^{E_7}\cdot L_{3}^{E_7}\cdot L_{corr}^{E_7}=$$
{\scriptsize
     \begin{multline*}
         \sinh\left[\frac{x}{4}: \right.\frac{(k+1)(k+2)^2(k+3)^3(k+4)^4(k+5)^5
     (k+6)^5(k+7)^4(k+8)^3(k+9)^2(k+10)(2k+7)(2k+8)}
     {1\cdot2^2\cdot3^3\cdot4^4\cdot 5^5\cdot 6^5\cdot 7^5\cdot 8^4}\times\\
     \frac{(2k+9)^2(2k+10)^2
     (2k+11)^3(2k+12)^2(2k+13)^2(2k+14)(2k+15)(3k+16)(3k+17)
     }{9^4\cdot 10^3 \cdot11^3\cdot12^2\cdot13^2\cdot14\cdot15\cdot16
     \cdot17}
     \end{multline*}     
 }
     
     which coincides with $D_Q^{E_7}$ of (\ref{D}).
     
     \subsection{$E_8$}
 For $E_8$ the Vogel parameters are $\alpha=-2, \beta=12, \gamma=20$.
$$L_{11}^{E_8}=\sinh\left[\frac{x}{4}: \right.
\frac{19\cdot 20\dots\ (k+18)}{1
\cdot1\dots k}=\sinh\left[\frac{x}{4}: \right.\frac{(k+1)(k+2)\dots(k+18)}{1\cdot2\dots18},$$
$$L_{12}^{E_8}=\sinh\left[\frac{x}{4}: \right.
\frac{18\cdot 19\dots\ (k+17)}{2
\cdot3\dots(k+1)}=\sinh\left[\frac{x}{4}: \right.
\frac{(k+2)(k+3)\dots(k+17)}{2\cdot3\dots17},$$
$$L_{13}^{E_8}=\sinh\left[\frac{x}{4}: \right.\frac{15\cdot 16\dots\ (k+14)}{5
\cdot6\dots(k+4)}=\sinh\left[\frac{x}{4}: \right.
\frac{(k+5)(k+6)\dots(k+14)}{5\cdot6\dots14}
,$$
$$L_{14}^{E_8}=\sinh\left[\frac{x}{4}: \right.
\frac{14\cdot 15\dots\ (k+13)}{6
\cdot7\dots(k+5)}=\sinh\left[\frac{x}{4}: \right.
\frac{(k+6)\dots(k+13)}{6\cdot7\dots13},$$
$$L_{15}^{E_8}=\sinh\left[\frac{x}{4}: \right.
\frac{11\cdot 12\dots\ (k+10)}{9
\cdot10\dots(k+8)}=\sinh\left[\frac{x}{4}: \right.\frac{(k+9)(k+10)}{9\cdot10}
,$$
$$L_{21}^{E_8}=\sinh\left[\frac{x}{4}: \right.
\frac{28\cdot 29\dots\ (2k+27)}{11
\cdot12\dots(2k+10)}=\sinh\left[\frac{x}{4}: \right.
\frac{(2k+11)\dots(2k+27)}{11\cdot12\dots27},$$
$$L_{22}^{E_8}=\sinh\left[\frac{x}{4}: \right.
\frac{24\cdot 25\dots\ (2k+23)}{15
\cdot16\dots(2k+14)}=\sinh\left[\frac{x}{4}: \right.
\frac{(2k+15)\dots(2k+23)}{15\dots23},$$
$$L_{23}^{E_8}=\sinh\left[\frac{x}{4}: \right.\frac{2k+19}{19}$$
$$L_{3}^{E_8}=\sinh\left[\frac{x}{4}: \right.\frac{(3k+28)(3k+29)}{28\cdot29},$$
$$L_{corr}^{E_8}=1.$$
The product of all these terms gives
$$X_2(x,k,-2,12,20) = L_{11}^{E_8}\cdot
     L_{12}^{E_8}\cdot L_{13}^{E_8}\cdot L_{14}^{E_8}\cdot
     L_{15}^{E_8}\cdot L_{21}^{E_8}\cdot L_{22}^{E_8}\cdot 
     L_{23}^{E_8}\cdot L_{3}^{E_8}\cdot L_{corr}^{E_8}=$$
     { \scriptsize
    
     \begin{multline*}
        \sinh\left[\frac{x}{4}: \right. \frac{(k+1)(k+2)^2(k+3)^2(k+4)^2(k+5)^3
     (k+6)^4(k+7)^4(k+8)^4(k+9)^5(k+10)^5(k+11)^4
     (k+12)^4}
     {1\cdot2^2\cdot3^2\cdot4^2\cdot 5^3\cdot 6^4\cdot 7^4\cdot
     8^4\cdot}\times\\
     \frac{(k+13)^4(k+14)^3(k+15)^2
     (k+16)^2(k+17)^2(k+18)(2k+11)(2k+12)(2k+13)(2k+14)(2k+15)^2
     }{9^5\cdot 10^5
     \cdot11^5\cdot12^5\cdot13^5\cdot14^4\cdot15^4\cdot16^4\cdot17^4}\times
     \\
     \frac{(2k+16)^2(2k+17)^2(2k+18)^2(2k+19)^3(2k+23)^2
     (2k+24)(2k+25)(2k+26)(2k+27)(3k+18)(3k+29)}{\cdot18^3\cdot19^3\cdot20^2\cdot21^2\cdot22^2\cdot23^2\cdot
     24\cdot25\cdot26\cdot27\cdot28\cdot29}.
     \end{multline*}
    }
coinciding with the direct calculation by (\ref{W}).

\end{document}